\documentclass{aa}  

\usepackage{graphicx}
\usepackage{float}
\usepackage{txfonts}
\usepackage{xcolor}

\newcommand{\HRIEUV}{HRI$_{\mathrm{EUV}}$\ }
\def\code#1{\texttt{#1}}
\begin{document}

   \title{Solar flares in the Solar Orbiter era: Short-exposure EUI/FSI observations of STIX flares}


   \author{Hannah Collier
          \inst{1,2}
          \and
          Laura A. Hayes \inst{3}
          \and
          Stefan Purkhart \inst{4}
          \and
          Säm Krucker \inst{1,5}
          \and
          Daniel F. Ryan \inst{1}
          \and
          Vanessa Polito \inst{6}
          \and
          Astrid M. Veronig \inst{4} 
          \and
          Louise K. Harra \inst{2}\fnmsep\inst{8}
          \and
          David Berghmans \inst{7}
          \and
          Emil Kraaikamp \inst{7}
          \and
          Marie Dominique \inst{7}
          \and
          Laurent R. Dolla \inst{7}
          \and
          Cis Verbeeck \inst{7}
          }
          % ...

   \institute{University of Applied Sciences and Arts Northwestern Switzerland (FHNW), Bahnhofstrasse 6, 5210 Windisch, Switzerland\\
              \email{hannah.collier@fhnw.ch}
        \and
        ETH Z\"{u}rich,
        R\"{a}mistrasse 101, 8092 Z\"{u}rich, Switzerland
        \and
        European Space Agency, ESTEC,
        Keplerlaan 1 - 2201 AZ, Noordwijk, The Netherlands
        \and
        Institute of Physics, University of Graz, Universitätsplatz 5, 8010 Graz, Austria
        \and
         Space Sciences Laboratory, University of California, 7 Gauss Way, 94720 Berkeley, USA
        \and
        Lockheed Martin Solar \& Astrophysics Laboratory, 3251 Hanover Street, Palo Alto, CA, 94304, USA
        \and
        Solar-Terrestrial Centre of Excellence (SIDC), Royal Observatory of Belgium, Ringlaan 3 Av. Circulaire, 1180 Brussels, Belgium
        \and
         PMOD/WRC, Dorfstrasse 33, CH-7260 Davos Dorf, Switzerland
        \\
         }

   \date{Received August 8 2024; accepted November 7 2024}

 
  \abstract
   {}
   {This paper aims to demonstrate the importance of short-exposure extreme ultraviolet (EUV) observations of solar flares in the study of particle acceleration, heating, and energy partition in flares. This work highlights the observations now available from the Extreme Ultraviolet Imager (EUI) instrument suite on board Solar Orbiter while operating in short-exposure mode.}
   {A selection of noteworthy flares observed simultaneously by the Spectrometer Telescope for Imaging X-rays (STIX) and the Full Sun Imager of EUI (EUI/FSI) are detailed. New insights are highlighted and potential avenues of investigation are demonstrated, including forward-modelling of the atmospheric response to a non-thermal beam of electrons using the RADYN 1D hydrodynamic code, in order to compare the predicted and observed EUV emission.}
   {The examples given in this work demonstrate that short-exposure EUI/FSI observations are providing important diagnostics during flares. A dataset of more than 9000 flares observed by STIX (from November 2022 until December 2023) with at least one short-exposure EUI/FSI 174 \AA\ image is currently available. The observations reveal that the brightest parts of short-exposure observations consist of substructure in flaring ribbons that spatially overlap with the hard X-ray emission observed by STIX in the majority of cases. We show that these observations provide an opportunity to further constrain the electron energy flux required for flare modelling, among other potential applications.}
   {}


   \maketitle
%

\section{Introduction}
During solar flares, the emission at ultraviolet (UV) and extreme ultraviolet (EUV) wavelengths can become locally enhanced by several orders of magnitude on a timescale of seconds to minutes. Even moderately sized C-class flares typically saturate current EUV and UV imagers as well as soft X-ray imagers. This means that spatial information in the saturated pixels is lost. Furthermore, depending on the type of detectors used, the total flux may or may not be conserved and neighbouring pixels might be affected (so called `bleeding'). Saturation during enhanced solar flare emission in such instruments is due to the fact that they are not exclusively designed to study flares. It is challenging to design an instrument with a dynamic range that is sufficiently large to observe both the faintest solar emission and the largest flares for a given exposure time. The saturation effect leads to challenges in many flare studies, particularly for relating coronal and chromospheric UV/EUV emission to observations from hard X-ray emission during the impulsive phase of flares.

One way to address the saturation issue of EUV imagers is to shorten the exposure time used. Some EUV/UV imagers, such as the Atmospheric Imaging Assembly (AIA) on board the Solar Dynamics Observatory, use automatic exposure control, in which flight software determines the exposure time based on the count flux \citep
{2012SoPh..275...17L}. However, with variations in this exposure time over the course of a flare, it can become a challenge to study given that the response of the detectors may be non-linear in exposure time, which can introduce artificial variations linked to the exposure time changes. Regardless, despite the variable exposure time setting, the exposure time is often too long and AIA observations of flares are frequently still saturated. Efforts have been made to overcome saturation issues through a reconstruction approach that utilises knowledge of the AIA diffraction pattern \citep{ raftery2011ApJ...743L..27R, krucker2011ApJ...734...34K, Schwartz_2014, Torre_2015InvPr..31i5006T, Guastavino_2019ApJ...882..109G, krucker2021ApJ...909...43K}. However, there are still challenges with total flux conservation and dealing with pixel bleeding. \cite{Kazachenko_2017ApJ...845...49K} also tackle this issue by linearly interpolating the flux pre- and post-saturation in order to infer spatial information about flare ribbons. An approach of this kind is, however, not suitable for photometric analysis.

In the era of multi-messenger observations, the availability of EUV imagers on the far side of the Sun from Earth has become increasingly important for event studies, since we cannot rely on Earth-based assets like AIA. For example, the Extreme Ultraviolet Imager (EUI) \citep{2020A&A...642A...8R, 2023A&A...675A.110B} on board Solar Orbiter \citep{2020A&A...642A...1M} provides EUV images for observations taken by the Solar Orbiter instrument suite as the spacecraft sweeps out an elliptical orbit, spending half of the mission duration on the far side of the Sun from Earth. EUI consists of three telescopes, the Full Sun Imager (EUI/FSI) and two High Resolution Imagers, which observe in Lyman $\alpha$, 174 \AA\ , and 304 \AA\ passbands. The focus of this work is on EUI/FSI observations.    

To combat the aforementioned saturation issues, EUI/FSI has been regularly taking short-exposure observations alongside normal-exposure frames in both the 174~\AA\ and 304~\AA\ channels since the end of 2022. This paper demonstrates the value of short-exposure EUV observations in flare studies and details the invaluable new datasets provided by EUI/FSI as well as recent observations with EUI's High Resolution Imager, EUI/\HRIEUV. In this work, we present a subset of interesting flares co-observed by the Spectrometer Telescope for Imaging X-rays (STIX) and discuss new insights enabled by this observation mode. 

\section{Observations}

\subsection{EUI/FSI and the short-exposure observing mode}
\begin{figure*}
     \centering
    \includegraphics[width=\textwidth]{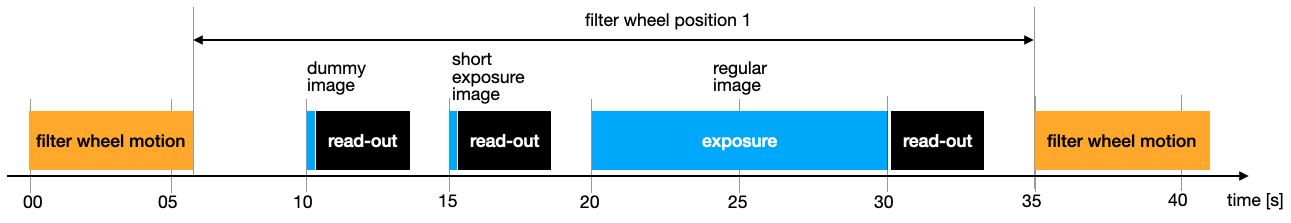}
    \caption{Typical time sequence of EUI/FSI synoptic imaging in between the filter wheel motions (orange boxes). Exposure times of the dummy image (0.2~s), the short-exposure image (0.2~s), and the regular image (10~s) are indicated in blue. Reading out a full sensor FSI frame (3072x3072 pixels) takes about 2.6~s seconds for any of the 3 exposures. The duration of the filter motions depends on the start and ending position (typically 3.7 - 6.2~s). During FSI synoptic imaging, this sequence is repeated for each of the two bandpasses with different filter positions, after which the filter wheel moves to a light-blocking position until, depending on the imaging cadence, a new cycle is started.}
    \label{fig:dutyCycle}
\end{figure*}

EUI/FSI is an EUV imager observing the full solar disc in the 174 \AA\ and 304 \AA\ EUV passbands. It continually provides synoptic observations with a variable cadence depending on telemetry allocations and observing mode. During FSI synoptic imaging, the two bandpasses are alternated by rotating a filter wheel \citep{2020A&A...642A...8R}.   Since November 9 2022, EUI/FSI has been regularly taking short-exposure observations alongside normal-exposure ones (see Fig \ref{fig:dutyCycle}). The nominal exposure time for an FSI image in either bandpass is 10~s. Reading out a full sensor FSI frame (3072x3072 pixels) takes about 2.6~s seconds.  Immediately before the regular exposure, two more images are taken during a very brief exposure time and read-out. The first of these is a dummy image that is disregarded upon arrival in the instrument computer. Taking a dummy image resets the sensor to a stable, reproducible status, independent from the previous history. The second extra image is the short-exposure image, which is intended to record intensity values where the (third) regular image is saturated.  We therefore chose a significantly shorter exposure time (1/50th, i.e. 0.2~s). Regular exposure FSI images contain intensity values in the range of 0–32767 DN. Unsaturated pixels (i.e. with values $\ll 32767$ DN) are not needed in the short-exposures as they are recorded in the regular image. Leaving a factor-two margin, we therefore limited the intensity range in the short exposures to values above $32767/(50\times2)$ DN = 327 DN.

The cadence of the EUI/FSI synoptics varies as a function of telemetry and can be between several minutes and half an hour. The observing settings described here are merely examples as the instrument is highly configurable and can be adjusted for different use-cases. The main constraint of deep space missions like Solar Orbiter is limited telemetry; however, as was discussed, for short-exposure images, only the brightest pixels, with pixel counts in the upper half of the dynamic range of regular images, are additionally downloaded. The other, dimmer and noisier pixels all get the same minimal value in the short-exposure image (327 DN), which facilitates compression of the short-exposure images greatly. Typically, the short-exposure data volume corresponds to $\sim16\%$ of the data volume of normal exposures for EUI/FSI images and as little as $\sim4\%$ for EUI/\HRIEUV frames.

\subsection{The STIX instrument}
STIX is a hard X-ray (HXR) imaging spectrometer measuring emission in the energy range 4-150~keV, from thermal and non-thermal flare accelerated electrons \citep{krucker2020_ins}. STIX observes the full solar disc at a minimum cadence of 0.5~s and operates continuously, unlike the other remote-sensing instruments on Solar Orbiter, which primarily operate during planned operational windows. STIX has been taking science data since early 2021, and to date (July 2024), STIX has observed more than 50,000 flares. It is an indirect Fourier imager, meaning that images are reconstructed from Fourier components sampled by the instrument (STIX samples 30 Fourier components or visibilities). There are certain constraints that arise from this fact, including a limited instrument dynamic range. This means that STIX is capable of only resolving the brightest X-ray emission. This limitation depends on counting statistics and relative brightness. For more information about the STIX imaging concept, we refer the reader to \cite{2023SoPh..298..114M}.

STIX provides us with critical diagnostics of solar flare energy release and heating through thermal and non-thermal HXR emission. Complementary EUV flare observations provide us with an insight into heated flare plasma with both high resolution and dynamic range. In particular, STIX observations give an insight into the released flare energy that is transferred to high-energy electrons, whereas EUI/FSI 304~\AA\ and 174~\AA\ give an insight into the chromospheric and coronal response to flare energy release, deposition, and heating. EUI provides snapshots of flares and flaring active regions that are critical for STIX observations, especially at times when Solar Orbiter is on the far side of the Sun from Earth, and therefore Earth-based context observations are not available. Additionally, short-exposure EUV observations help to verify the sources reconstructed with STIX imaging algorithms such as \texttt{Clean} \citep{hurford2002SoPh..210...61H}, \texttt{MEM\_GE} \citep{Massa_2020}, and \texttt{forward-fit PSO} \citep{2022A&A...668A.145V}, the presence of which may otherwise be inconclusive due to dynamic range limitations. Furthermore, the highly resolved flare kernels obtained from EUV flare observations provide important constraints on flare energy deposition rates.

\section{Available data and example case studies}
A common flare list has been created of STIX flares with short-exposure EUI/FSI observations. This list was compiled from the operational STIX flare list available at the \texttt{STIX data center}\footnote{https://datacenter.stix.i4ds.net/} and is also accessible via the API of the Python library \code{stixdcpy}\footnote{https://github.com/i4Ds/stixdcpy} \citep{hualin_stix}. Flares with > 1000 peak counts in the 10-15~keV quicklook channel were considered. The available Level 2 (L2) EUI/FSI 174~\AA\ short-exposure files within the flare time range were obtained through the SIDC website\footnote{https://www.sidc.be/EUI/data/, these data are also available from https://soar.esac.esa.int/soar/}. From this, a combined list of STIX-EUI/FSI short-exposure observations was created. This list considers the time range from November 9 2022 to November 31 2023 (the date from which short-exposure observations began until the most recent publicly available L2 EUI files at the time of analysis). We find that $\sim 42$ \% of STIX flares in this time range have at least one short-exposure frame, which amounts to 9481 co-observed flares; the code used to create this list is publicly available\footnote{https://github.com/hayesla/eui\_gi\_visit}. From this flare list, a subset of events have been further analysed and three events that demonstrate the advantages of this unique dataset are presented in the sections that follow. 

\subsection{STX2023-07-16T04:32: Rapid pulsations}
\begin{figure*}
     \centering
    \includegraphics[width=0.9\textwidth]{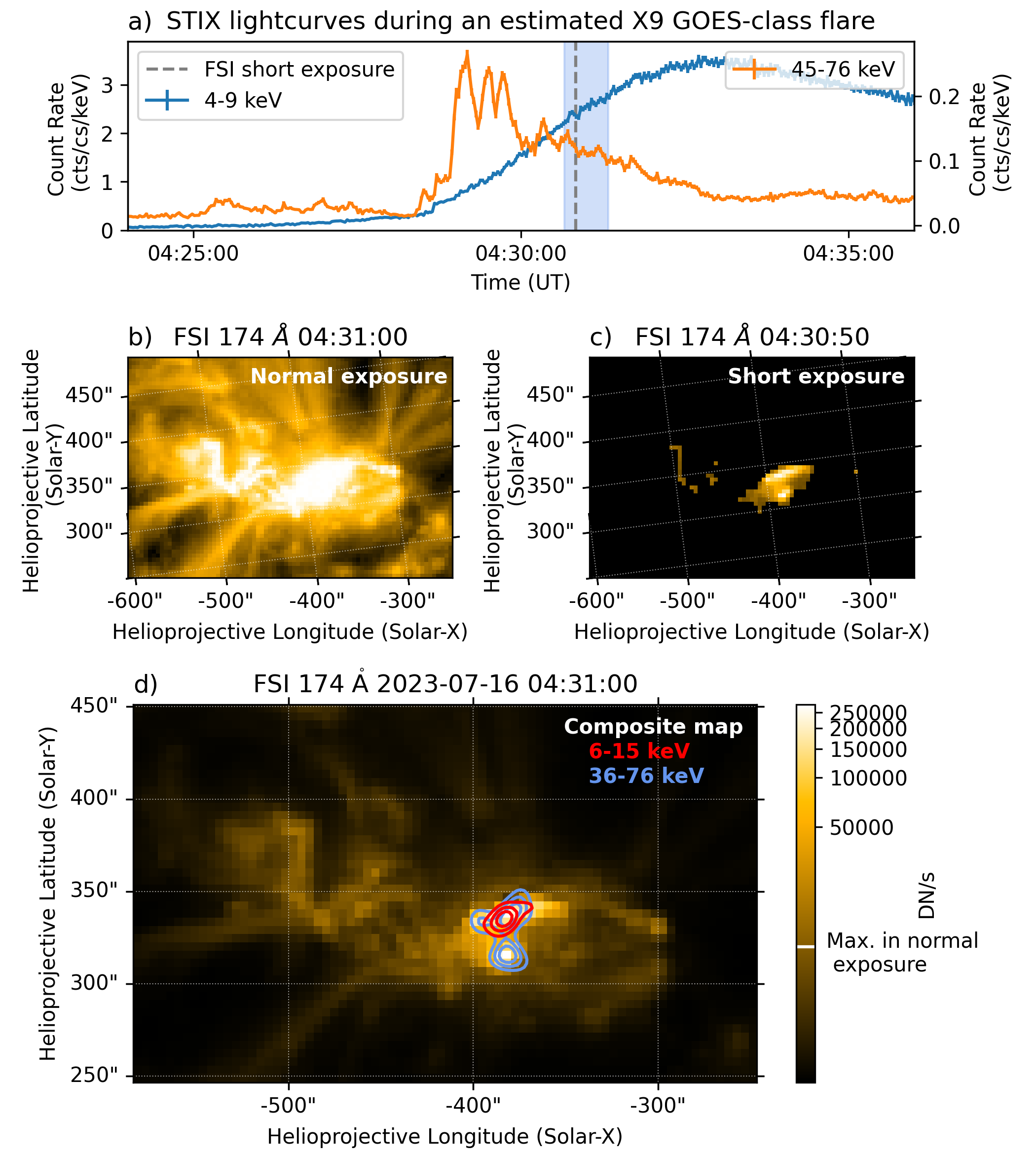}
    \caption{Overview of the STX2023-07-16T04:32 (estimated X9 GOES class) flare. Panel a) shows the STIX light curves. The non-thermal emission (45-76 keV) shows rapid variation on a timescale of $\sim 10$~s. The thermal (4-9 keV) light curve is shown for the background detector only, as this detector is unaffected by attenuator motion (the attenuator was inserted during this flare). Panels b) and c) show the normal- and short-exposure observations of the flare site, respectively. Panel d) is a composite map with thermal (6-15 keV, in red) and non-thermal HXR contours (36-76 keV, in blue). The 40, 60, \& 80\% contour levels are shown.  The fine structure revealed by the EUI/FSI short-exposure frame corresponds well to the HXR emission observed by STIX. This map has been rotated such that Solar north is pointing upwards, unlike panels b) and c). The maximum flux in the normal exposure frame is highlighted in the colour bar of panel d), highlighting the gain in dynamic range provided by the short-exposure observation.}
    \label{fig:160723_overview}
\end{figure*}

STX2023-07-16T04:32\footnote{The STX flare identifier used throughout this work labels each flare by the date and time at which the flare counts reached a maximum in the STIX 4-10 keV quicklook channel.} is one of the largest flares observed by STIX to date. It was estimated to be of GOES class X9 using the method proposed by \cite{hualin_stix}. At the time, Solar Orbiter was at $157^{\circ}$ to the Sun-Earth line and as the flare was located at $\sim(135, 19)^{\circ}$, in heliographic stonyhurst coordinates, the flare was not observable from Earth. The flare displayed rapid variation on timescales of $\sim$ 10~s in the non-thermal HXR emission (see Fig. \ref{fig:160723_overview}, panel a), making it an interesting case in which to study quasi-periodic pulsations and rapid variation in flaring emission \citep[e.g.][]{collier_2023, collier24}. EUI/FSI observed this flare in both short- and normal-exposure mode after the non-thermal peak time but before the thermal peak. The time of the short exposure is denoted by the vertical dashed line in Fig. \ref{fig:160723_overview} a). In the normal exposure frame (panel b), the flare site is saturated; however, the short exposure (panel c) distinctly shows the flaring ribbons and the fine structure within; that is, the short exposure frame highlights the distribution of brightness along the length of the northern ribbon as well as the spatially compact bright source in the southern flare ribbon. Due to a known non-linearity in the instrumental response with exposure time, a correction factor is required to match the fluxes observed by the normal and short exposures (E. Kraaikamp 2024, priv. communication). The cause of the non-linearity is currently being investigated. In this case, the flux correction factor used was 1.33. This factor was obtained by matching the flux in non-saturated pixels in the regular exposures with the corresponding pixels in the short exposures for a selection of flares. It is therefore effectively a shortening of the exposure time by this factor. Given the current level of understanding of this discrepancy, it is recommended that this analysis be performed on a case-by-case basis. Panel d) shows a composite map of the short and normal exposure frames. It has been rotated such that Solar north is up for comparison with the STIX contours. Since the counts in the short-exposure maps are spatially sparse, performing accurate rotations can be tricky as this typically involves interpolation between pixels. In this work, the rotation functionality provided by \code{sunpy} \citep{sunpy_community2020}, \code{sunpy.map.GenericMap.rotate}, has been used. In particular, the `scikit-image'  method was used as the default rotation method (`scipy') uses spline interpolation, which involves a pre-filtering step, and this step causes discontinuities to have ripple-like effects. This effect becomes particularly prominent with short-exposure maps as there are sharp spatial discontinuities in flux. In panel d) the non-thermal (36-76 keV) and thermal (6-15 keV) HXR emission is shown for the time range 04:30:40-04:31:20 UT -- shaded in blue in panel a) (all times given in this work are at the observing telescope). A shift of (-10, 35)'' has been applied to align the STIX maps with the short-exposure frame. Interestingly, the separation of the HXR non-thermal sources corresponds remarkably well to that of the ribbons revealed by the short-exposure EUI/FSI frame. The thermal loop appears to sit on the northern ribbon; however, this is likely due to projection effects -- the flare was located at approximately $135^{\circ}$ from the central meridian, while Solar Orbiter was at $157^{\circ}$. The 45-76 keV light curve of this flare shows rapid temporal variation. This temporal variability is a signature of fundamental energy release and particle acceleration processes ongoing in flares, and therefore is a crucial diagnostic required to reach a unified flare model. The spatial evolution of the atmospheric response to this electron beam variability is currently unresolved temporally, given the low-cadence EUI/FSI observations while in synoptic mode. While these synoptic observations are vitally important for understanding the surrounding active region structure, the need for high-time-cadence short-exposure observations of flaring ribbons in EUV is apparent.     

\subsection{STX2022-11-13T06:18: A `standard' flare}
\begin{figure*}
    \centering
    \includegraphics[width=0.85\textwidth]{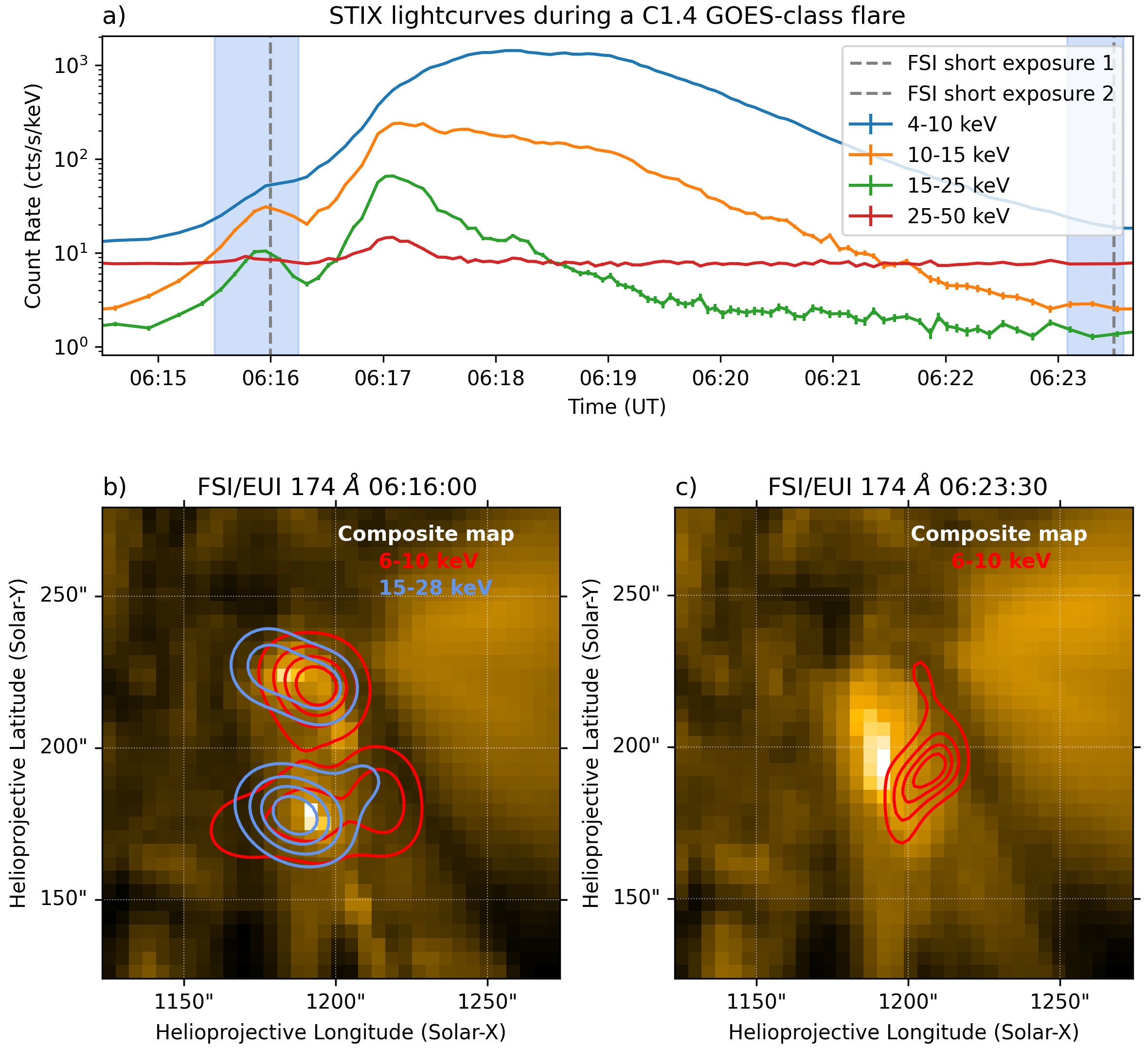}
    \caption{Overview of STX2022-11-13T06:18. Panel a) shows the STIX light curves. The times of EUI/FSI short-exposure frames are denoted by dashed lines, and the time intervals used to create the STIX images shown in panels b) and c) are shaded in blue. Panels b) \& c) show FSI/EUI 174 \AA\ composite maps of the short- and normal-exposure exposure observations for the early and late frame, respectively. Panels b) and c) also show the STIX thermal (6-10 keV) and non-thermal (15-28 keV) emission overlaid on the EUI/FSI short-exposure frames for the two times, respectively. The 20, 40, 60, and 80\% contour levels are shown. The maps shown have been rotated such that Solar north is pointing upwards.}
    \label{fig:13-11-22}
\end{figure*}

STX2022-11-13T06:18 was a C1.4 GOES-class flare observed by both Solar Orbiter and Earth-based observatories. At this time, Solar Orbiter was at a distance of 0.64~AU from the Sun and an angle of $20.7^\circ$ to the Sun-Earth line. Two EUI/FSI short-exposure frames were obtained during the flare, one during the first non-thermal HXR burst and the second towards the end of the flare (see Fig. \ref{fig:13-11-22} a). Figures \ref{fig:13-11-22} b) and c) are composite maps of the normal- and short-exposure EUI/FSI frames from the times denoted by the vertical dashed lines in panel a). In such composite maps, all non-zero short-exposure pixel values have replaced their corresponding normal exposure pixel counts. The counts were normalised by exposure time and the flux correction factor was applied as for STX2023-07-16T04:32. The corresponding AIA 171 \AA\ observations are shown in Fig. \ref{fig:131122_aia_frames}.

This flare is another example of a flare whose signatures are consistent with the standard flare `CSHKP' model \citep{carmichael_1964NASSP..50..451C, sturrock_1966Natur.211..695S, hirayam_1974SoPh...34..323H, kopp_1976SoPh...50...85K}. During the first HXR burst, we see non-thermal emission at 15-28 keV from two footpoints and thermal emission from the base of the flare loop emitted at 6-10 keV (Fig. \ref{fig:13-11-22} b). The time interval used to reconstruct the STIX images in Fig. \ref{fig:13-11-22} b) was 06:15:30-06:16:15 UT. Interestingly, two distinct brightenings are seen in EUI/FSI 174~\AA\ , which align well with the HXR footpoints. Later in the flare, the previously heated thermal loop connecting the two footpoints is observable in the 174~\AA\ EUI/FSI channel after cooling to temperatures the channel is most sensitive to (the temperature response peaks at $\sim$1~MK). The EUI short-exposure frame shows a low-lying or tilted loop formed in the early phase of the flare and extending between the two footpoints observed in the previous frame. In contrast, the 6-10~keV emission (for the interval 06:23:05-06:23:35 UT) shows the hotter, higher, and more arch-shaped loops formed during the later phase of the flare. 

This example demonstrates the need for high-cadence short-exposure observations. The EUI/FSI channels are sensitive to a range of temperatures, and therefore see both the impulsive impact of non-thermal particles as well as more gradual heating in flares. The short-exposure EUI/FSI frame enables us to identify the brightest of the saturated pixels, and hence enables us to confidently align the STIX non-thermal sources. Fortunately, in this case there was a short-exposure frame that captured the heating of flare ribbons caused by the impact of non-thermal particles as well as post-flare heating or cooling. However, since the response of the 174~\AA\ channel is temperature-sensitive, and thus changes significantly in time, in order to investigate the exact correspondence between EUV and HXR emission in flares, high-time-cadence observations are required.  

In order to verify our understanding of how the emission observed in the 174 \AA\ passband short-exposure flare observations relates to the injected electron beam, we simulated the atmospheric response to an injected electron beam, the parameters of which are inferred from HXR measurement. Furthermore, as this flare is a classic example of the standard flare model at work, it lends itself well to modelling efforts. We did this using the 1D radiative hydrodynamic RADYN code \citep{carlson1992ApJ...397L..59C, carlson1995ApJ...440L..29C, carlson1997ApJ...481..500C, carlson2002ApJ...572..626C, abbett1999ApJ...521..906A, 2005ApJ...630..573A, 2015ApJ...809..104A}. The input electron beam parameters were derived by fitting the HXR spectrum during the first non-thermal burst (06:15:29-06:16:16 UT) with a thermal model (\texttt{f\_vth.pro}), a thick target bremsstrahlung component (\texttt{f\_thick2.pro}), and an albedo component using the \texttt{OSPEX} spectral fitting software. The data between 4-36 keV was included in the fitting procedure and the resultant fit obtained a reduced chi-square, $\chi^2_{\nu} = 0.87$, assuming 5\% systematic errors. The fitted models indicated a non-thermal electron beam with an electron rate of $\alpha = (0.34 \pm 0.04) \times 10^{35}$ electrons/s, a spectral index of $\delta = 4.97 \pm 0.09$, and a low energy cut-off of $E_c = 13.37 \pm 0.57$ keV, from which we obtained a total non-thermal power of $9.72 \times 10^{26}$ erg/s. The power obtained is an estimate of the total spatially integrated power in the electron beam. In order to estimate the area over which this electron beam is distributed, we calculated the footpoint area. In this case, the advantage of having unsaturated footpoint observations was weakened by the fact that the spatial resolution of EUI/FSI at this radial distance is lower than that of AIA. Therefore, estimating the footpoint area using the EUI/FSI short-exposure frame gives a similar result to AIA. We estimated the footpoint area by calculating the area of the pixels within the 30\% AIA contour level during the first non-thermal burst. The choice of the contour level is somewhat arbitrary; however, in this case 30\% was chosen as a middle ground between flux over- and underestimation. From this, we obtained a total footpoint area estimate of $A \approx 10^{17} {\mathrm{cm}}^2$. This results in a total electron beam energy flux of $9 \times 10^9$ erg/s/$cm^2$, which is used as an input in the RADYN model. Furthermore, the loop length was estimated by simply using half the distance between footpoint brightenings in AIA 171 \AA\ as an estimate of the radius of a semi-circular loop, from which a half-loop length (from loop apex to footpoint) of approximately 17 Mm was obtained. 

An important consideration is the impact of the choice of starting atmosphere on the model results. In the RADYN run presented here, a starting apex temperature of 3 MK and a simple starting atmosphere (VAL3C), similar to what is used in the FCHROMA grid database \citep{FCHROMA_2023A&A...673A.150C}, was used. \cite{polito2018ApJ...856..178P} have demonstrated the impact of initial loop-top temperatures for nanoflare emission. In particular, \cite{polito2018ApJ...856..178P}
show that for higher starting temperatures, electrons are typically stopped at high altitudes in the corona, and therefore only a small amount of energy is dissipated in the transition region and chromosphere. This has a direct impact on the intensity of EUV emission.

A triangular beam of duration 45 s with the aforementioned energy flux, a low energy cut-off of 13~keV, and spectral index of 5 were input in the model and the 1D atmospheric response was simulated. The model outputs include a grid of temperature and ambient ionised hydrogen density as a function of height along the loop as a function of time. The emission measure and temperature in each grid cell was then folded through the latest EUI/FSI 174 \AA\ response (2024, priv. comm. with Frédéric Auchere) in order to determine the predicted response. Since this wavelength range is optically thin, the total response can be obtained by integrating along the entire loop length. 

\begin{figure}
    \centering
    \includegraphics[width=0.9\columnwidth]{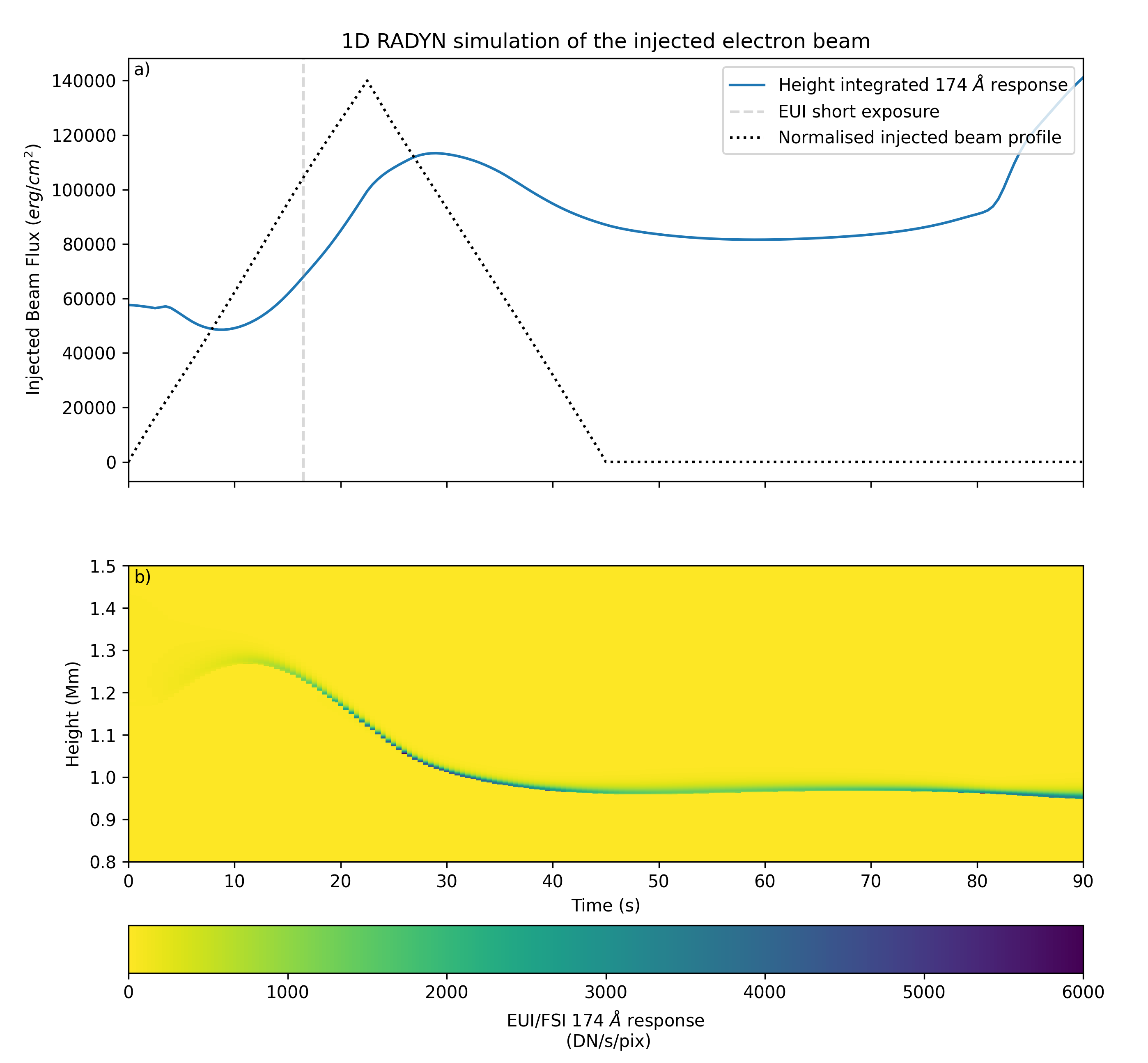}
    \caption{Predicted EUI/FSI 174 \AA\ response to a non-thermal beam of electrons whose impact on the solar atmosphere was modelled by 1D RADYN simulations. Panel a) shows the predicted EUI/FSI 174 \AA\ flux integrated over the entire loop. At the time of the short-exposure observations, the integrated 174 \AA\ flux is $\sim 68,000$ DN/s/pix, which is remarkably consistent with the peak rate observed, which reaches $\sim 74,000$ DN/s/pix.  Panel b) also shows the response as a function of height in the chromosphere, between 0.8 and 1.5 Mm above the photosphere. The data has been clipped to a maximum of 6000 DN/s/pix in order to increase image contrast for display purposes.
 }
    \label{fig:radyn_results}
\end{figure}

It is important to note that the simulated grid is adaptive, which means that the cell boundaries vary in time in order to enable accurate simulations of shocks and sharp discontinuities. For representation purposes, Fig. \ref{fig:radyn_results} b) was obtained by interpolating the temperature and density grids to a fine (0.0003 Mm resolution), evenly sampled grid and folding the resampled grid through the EUI/FSI 174 \AA\ response function. This introduces minor errors due to interpolation; however, for the purpose of demonstrating the height of the observed emission, the interpolation accuracy is sufficient.

Figure \ref{fig:radyn_results} a) shows the predicted EUI/174 \AA\ flux given the simulated atmosphere, integrated in height. The EUI/FSI short-exposure frame was taken approximately 16.5 s into the beam injection. At this time, the total predicted EUI/FSI 174 \AA\ flux is $\sim 68,000$ DN/s/pix, which agrees remarkably well with the observed peak flux in the short-exposure frame whose maximum is $\sim 74,000$ DN/s/pix. Figure \ref{fig:radyn_results} b) shows the predicted 174 \AA\ flux as a function of height and time. As was expected, the majority of the flux detected originates from a very thin layer in the chromosphere (0.9-1.3 Mm above the photosphere). This indicates that the majority of the 174~\AA\ EUV flux during the impulsive phase of the flare originated from the chromosphere. In comparison, coronal emission is very faint due to both the significantly higher temperatures and lower emission measure at this time. 

\subsection{STX2023-04-22T22:21: Potential anchor points of an erupting filament}
\begin{figure*}
    \centering
    \includegraphics[width=\textwidth]{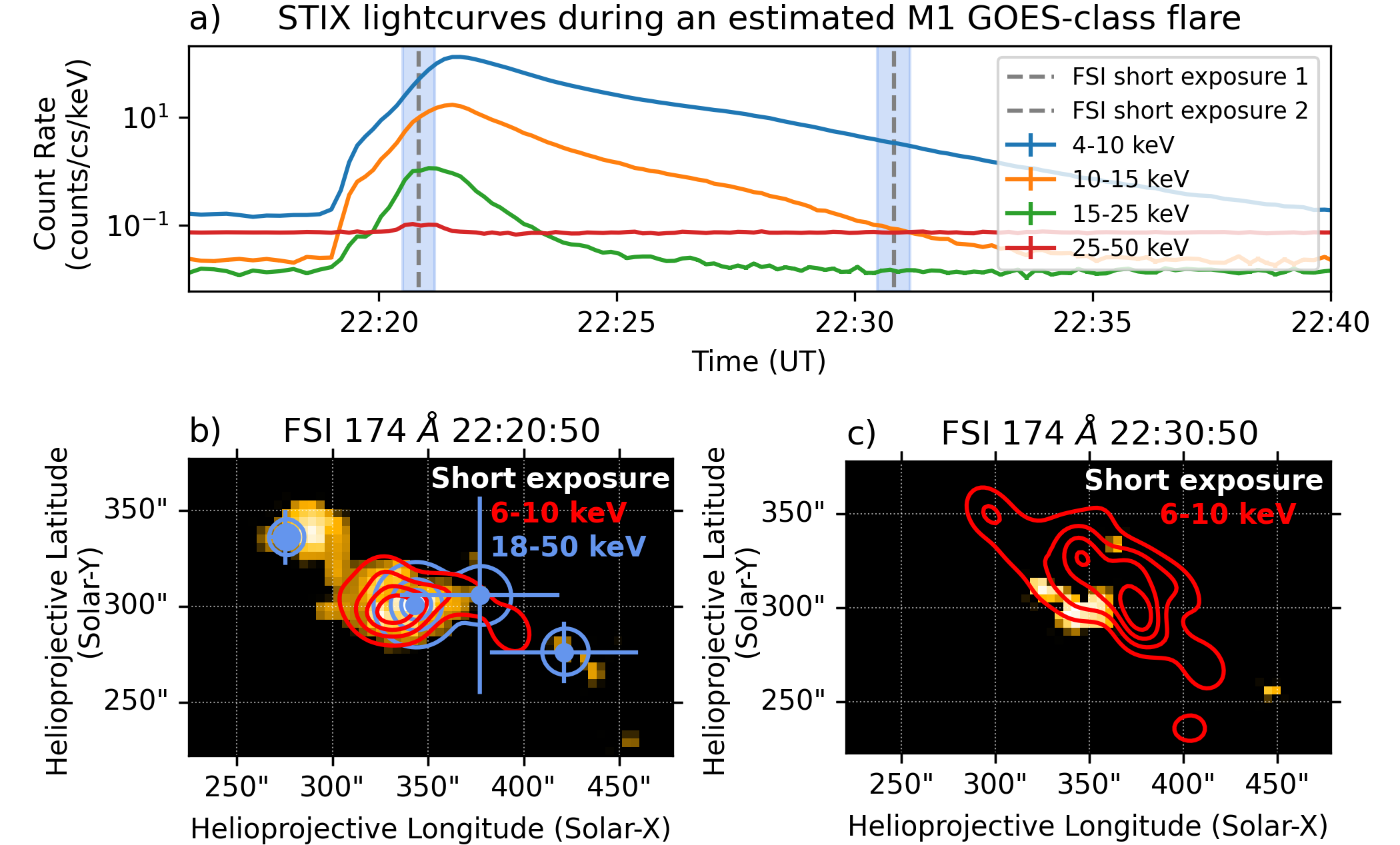}
    \caption{Overview of STX2023-04-22T22:21 showing STIX light curves (panel a) and two EUI/FSI short-exposure frames (panels b and c). The first short-exposure frame shows four non-thermal sources both in HXR (blue contours, 18-50 keV) and EUV 174~\AA\ emission. The red contours show the 6-10 keV thermal HXR emission. We interpret that emission in the later EUI frame as the previously heated flare loop that has cooled sufficiently such that it was observable in the 174~\AA\ passband. The thermal emission from STIX at this time originates from an elongated source that only appears in the EUI/FSI short-exposure frame at 23:00:50 (not shown). As before, the maps shown here have been rotated such that north is up.}
    \label{fig:22-04-23}
\end{figure*}

STX2023-04-22T22:21 was an estimated M1 GOES-class flare \citep{hualin_stix}. It occurred while Solar Orbiter was at a distance of 0.4 AU from the Sun and an angle of $129^{\circ}$ from the Sun-Earth line. The flare was not observed by Earth-based instruments. Figure \ref{fig:22-04-23} a) shows the STIX light curves with the time of the short-exposure observations denoted by the vertical dashed lines. Panels b) and c) show the EUI/FSI short-exposure observations at these times with STIX contours overlaid. The blue contours in b) are the 5-85\% contours levels of the 18-50 keV emission during this time. The map has been created by forward-fitting four circular Gaussian sources with all other parameters (location, source size, and flux) free  for the time range 22:20:30-22:21:10 UT, using the method detailed by \cite{2022A&A...668A.145V}. The position and error on the coordinates are shown. The red contours in panels b) and c) denote the 20-80\% contours levels of the 6-10 keV emission of both 40\,s intervals centred around the EUI/FSI short-exposure time. In panel c), there is emission from previously heated loops that have presumably cooled to temperatures that the EUV passband is highly sensitive to. The thermal HXR emission at this time originates from an elongated source that spans the extent of the original non-thermal structure shown in panel b). 

The four non-thermal sources of HXR emission are of interest here as they are thought to correspond to the anchor points of an erupting flux rope. \cite{2023A&A...670A..89S} recently observed non-thermal HXR emission from the footpoints of an erupting filament alongside the standard flare footpoint emission. Similar observations have been reported at microwave wavelengths by \cite{Chen_2020}, who observed microwave emission coming from the filament legs during the impulsive phase of the flare. In both cases, the emission was interpreted as a signature of non-thermal electrons that propagate along the flux rope post-reconnection and finally reach chromospheric altitudes, resulting in the observed HXR and microwave emission, respectively. Such observations can be used to study transport effects on particles in an erupting flux rope. Furthermore, such observations can shed light on the acceleration process of electrons in flares.

As Earth observations of this flare are not available, EUI/FSI observations are essential for providing context to STIX observations. This is particularly important here as four sources are resolved with STIX, the faintest of which is only at the 5\% contour level. This observation is therefore pushing the boundary of what STIX is capable of resolving given its limited dynamic range. However, given the strong correspondence with the brightest EUV sources seen in the short-exposure frame, we are confident that STIX is reliably resolving four real non-thermal sources. 

\section{Discussion and summary}

\subsection{The importance of the EUI/FSI short-exposure observing mode}
Short-exposure EUV/UV observations are a key observational asset for flare studies and are particularly important for missions like Solar Orbiter that frequently observe the Sun from the `back-side' and that therefore do not have context observations from Earth-based assets. Short-exposure observations of flares are important for precisely identifying flare locations. Specifically, EUI/FSI short-exposure observations can be utilised to correct the absolute pointing of STIX, which is a significant advantage of having the EUI and STIX instruments on the same spacecraft. These observations are also useful for identifying fine structure within flaring ribbons, active regions, and other transients such as coronal mass ejections. EUI/FSI on board Solar Orbiter has been routinely collecting such data since November 2022, with more than 9000 flares co-observed by STIX and EUI/FSI from then until December 2023. 

\subsection{Extending the short-exposure mode to EUI/\HRIEUV}
While EUI/FSI observations are critical, what is really needed in order to advance our understanding of particle acceleration, energy release, and transport in flares is high-cadence short-exposure EUV observations with high spatial resolution. Very high-cadence ($\approx$0.3-1.7s) images and spectral observations from the Interface Region Imaging Spectrograph \citep[IRIS;][]{DePontieu2014,DePontieu2021} have  provided significant new insights into plasma dynamics occurring in flares, which cannot be observed at lower cadence \citep{Jeffrey18,French21,Lorincik22}. Furthermore, with high-cadence, short-exposure observations the spatial origin of rapid variations in the observed emission could be determined, a pursuit that is particularly important for the study of quasi-periodic pulsations in flares \citep[e.g.][]{zimovets2021}. Resolving the spatial origin of such variation is a key component in understanding particle acceleration in flares (\cite[e.g.][]{collier24}; see also discussion in \cite{2023BAAS...55c.181I}). These recent studies have motivated the need for regular, unsaturated, and high-cadence flare observations.

In order to do this, co-ordinated efforts are under way in which Solar Orbiter instruments are operating with flare-optimised settings during remote sensing windows, with support from Earth-based assets. In such modes, the suite of relevant remote sensing  instruments on board Solar Orbiter are pointed at the active region predicted to be the most likely to flare. During the recent  perihelion of March and April 2024, many C-and several M- GOES class flares were observed during this flare campaign with completely unprecedented resolution. For example, the M-class flare of March 19 2024 was observed while EUI/\HRIEUV was operating at high cadence (2\,s), taking both normal- (2\,s) and short- (0.04\,s) exposure observations with a 2-pixel resolution of 310 km. Such observations promise enlightening insights into particle acceleration, heating, and energy partition in solar flares. A follow-on paper will discuss the high-cadence, short-exposure EUI/\HRIEUV data obtained during the March-April 2024 flare campaign in more detail. Another major flare campaign is planned for the perihelion of Solar Orbiter in March-April 2025. 

\subsection{Potential new science}
Short-exposure EUV/UV observations provide complementary diagnostics of the flaring region. ‘Far-side’ context observations are becoming increasingly pertinent in the era of multi-messenger solar physics. In this work, we have highlighted the invaluable data that the EUI/FSI instrument on board Solar Orbiter is now routinely obtaining. These new datasets reveal fine structure in flaring ribbons that is typically lost due to instrument saturation.

Importantly, from these observations, better constraints on the energy flux of non-thermal electron beams, required as input for flare modelling, can be derived, particularly when Solar Orbiter is at a close radial distance to the Sun. This will be especially revolutionary given the unprecedented spatial resolution of EUI/\HRIEUV (200 km 2-pixel resolution on the surface of the Sun) at perihelion \citep{2023A&A...675A.110B}. By combining unsaturated high-cadence observations of flaring ribbons from HRI$_{EUV}$ and STIX on board Solar Orbiter, the non-thermal electron energy flux in flares can be determined much more precisely than was previously possible. Finally, high-cadence unsaturated observations of flaring loops will enable us to study flare heating and cooling processes in more detail.  

In summary, the need for unsaturated EUV/UV observations of solar flares has been demonstrated in this paper. Currently, EUI on board Solar Orbiter is providing excellent observations of this kind. Looking to the future, we note that there is a strong need for a high-cadence flare-focused EUV/UV imager for the next solar cycle \citep{2023BAAS...55c.181I}. 

\begin{acknowledgements}
Solar Orbiter is a space mission of international collaboration between ESA and NASA, operated by ESA. The STIX instrument is an international collaboration between Switzerland, Poland, France, Czech Republic, Germany, Austria, Ireland, and Italy.     
HC and SK are supported by the Swiss National Science Foundation Grant 200021L\_189180 for STIX.
L.A.H is supported by an ESA Research Fellowship. 
This research was funded in part by the Austrian Science Fund (FWF) 10.55776/I4555.
This work was additionally supported by the EUI PRODEX Guest Investigator Programme. The EUI instrument was built by CSL, IAS, MPS, MSSL/UCL, PMOD/WRC, ROB, LCF/IO with funding from the Belgian Federal Science Policy Office (BELSPO/PRODEX PEA 4000134088, 4000112292, 4000136424, and 4000134474). E.K.\, C.V.\, M.D.\ and L.R.D.\ thank the Belgian Federal Science Policy Office (BELSPO) for the provision of financial support in the framework of the PRODEX Programme of the European Space Agency (ESA) under contract numbers 4000143743, 4000134088, 4000134474 and 4000136424. This manuscript made use of several open source packages including version 5.0.0 (https://doi.org/10.5281/zenodo.8037332) of the SunPy open source software package \citep{sunpy_community2020} and version 0.1.2 of the stixpy package (https://github.com/TCDSolar/stixpy) . It also used version v0.5.2 of the IDL STIX software (https://github.com/i4Ds/STIX-GSW).

\end{acknowledgements}
\bibliographystyle{aa} 
\bibliography{aanda.bib}

\begin{appendix}
\onecolumn
\section{STX2022-11-13T06:18: AIA observations} \label{appendix:aia_obs}
\begin{figure}[H]
    \centering
    \includegraphics[width=\textwidth]{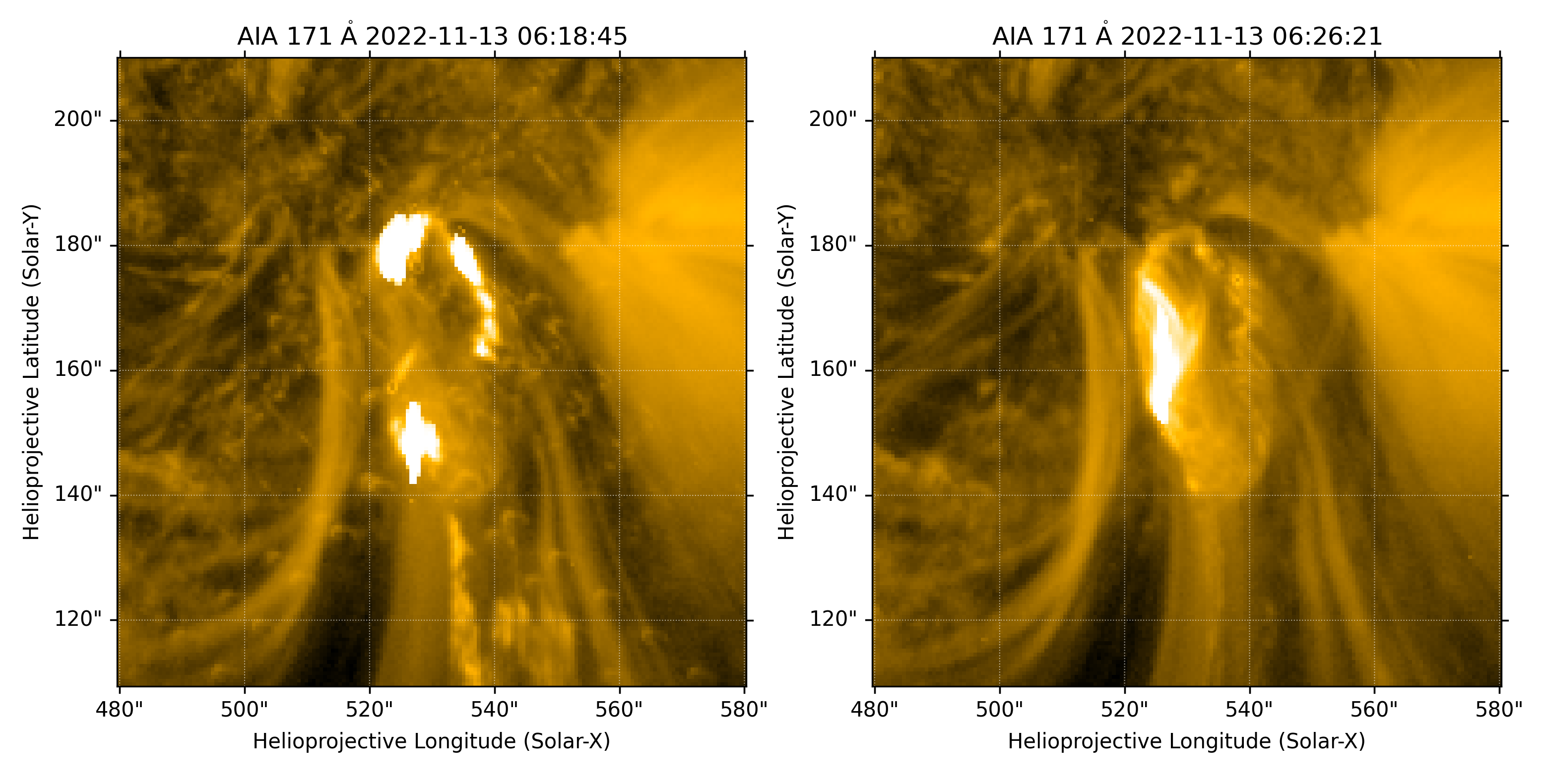}
    \caption{AIA 171 \AA\ observations closest in time to the EUI/FSI short-exposure frames from the STX2022-11-13T06:18 flare. The times shown correspond to panels b) and c) of Fig. \ref{fig:13-11-22}. The flaring pixels are clearly saturated in both frames. The footpoints in the early frame also suffer from the effect of pixel bleeding.
    }
    \label{fig:131122_aia_frames}
\end{figure}
 
\end{appendix}

\end{document}